\documentclass[conference,10pt]{IEEEtran}
\usepackage{float}
\usepackage{cite}										
\usepackage{amsmath}									
\usepackage{graphicx}
\usepackage{epstopdf}
\usepackage{float}
\usepackage{bm,upgreek}
\emergencystretch=\maxdimen
\usepackage{amsfonts}
\usepackage{algorithm}
\usepackage{algpseudocode}

\usepackage{enumitem}
\allowdisplaybreaks
\ifCLASSINFOpdf
\else

\fi

\begin{document}
\title{Distributed Gauss-Newton Method for AC State Estimation: A Belief Propagation Approach}

\author{\IEEEauthorblockN{Mirsad Cosovic}
\IEEEauthorblockA{Schneider Electric DMS NS LLC\\
Novi Sad, Serbia\\
Email: mirsad.cosovic@schneider-electric-dms.com}
\and
\IEEEauthorblockN{Dejan Vukobratovic}
\IEEEauthorblockA{Department of Power, Electronics and Communications Engineering, \\
University of Novi Sad, Serbia\\
Email: dejanv@uns.ac.rs}}

\maketitle

\begin{abstract}
In this paper, we propose a solution to an AC state estimation problem in electric power systems using a fully distributed Gauss-Newton method. The proposed method is placed within the context of factor graphs and belief propagation algorithms and closed-form expressions for belief propagation messages exchanged along the factor graph are derived. The obtained algorithm provides the same solution as the conventional weighted least-squares state estimation. Using a simple example, we provide a step-by-step presentation of the proposed algorithm. Finally, we discuss the convergence behaviour using the IEEE 14 bus test case. 
\end{abstract}


\IEEEpeerreviewmaketitle

\section{Introduction}
The state estimation (SE) is an essential part of the real-time energy management system (EMS), and it provides inputs for other EMS functions. A SE algorithm, jointly with network topology processors, observability analysis and bad data analysis, provides an estimate of the system state according to the network topology and available measurements. SE is performed on a bus/branch model and used to reconstruct the state of the observable part of the system. Conventional SE algorithms are centralized and typically use the Gauss-Newton method to solve the non-linear weighted least-squares (WLS) problem \cite{monticelli}, \cite{abur}.

Deregulation of electric power systems implies their decentralized structure, however, integrated control and monitoring across the entire network is still needed. In view of recent trends, control centers are mostly migrating toward distributed control centers \cite{wu}. Consequently, many centralized algorithms of EMS have to be redefined, requirements being distributed and computationally more efficient algorithms. 

Probabilistic graphical models, such as factor graphs (FGs), represent a powerful tool for modeling probabilistic systems. The algorithm for exact inference on probabilistic graphical models without loops is known as the belief propagation (BP) algorithm \cite{pearl}, \cite{bishop}. Using BP, it is possible to efficiently calculate marginal distributions or a mode of the joint distribution of the system of random variables. The BP algorithm can be also applied to graphical models with loops (loopy BP)\cite{loop}, although in that case, the solution is not guaranteed to converge to the correct marginals/modes of the joint distribution. BP is a fully distributed algorithm that takes probability distributions as an input, processes them, and outputs marginal probability distributions used to estimate values of state variables. This makes it a flexible solution for accommodation of distributed power sources and time-varying loads in various applications of electric power systems.

The work in \cite{kavcic} provides the first demonstration of BP applied to the SE problem. Although this work is elaborate in terms of using, e.g., environmental correlation via historical data, it applies BP to a simple linearized DC model. The AC model is recently addressed in \cite{ilic}, where tree-reweighted BP is applied using preprocessed weights obtained by randomly sampling the space of spanning trees. 

In this paper, we also solve the AC model via BP but in a completely different framework that we find simpler and more intuitive. We consider the AC SE model that we cast into a FG representation and solve using the BP algorithm. The proposed BP algorithm is obtained after a transformation of the initial WLS problem into the maximum a posteriori probability (MAP) problem that estimates the vector of increments of the state variables. The resulting BP algorithm has the interpretation of a fully distributed Gauss-Newton method with the same accuracy as the conventional or centralized SE. Consequently, the presented algorithm applies similar mathematical framework as the conventional SE, which simplifies its integration into existing EMS. In addition, it can be implemented in a fully distributed manner suitable for the distributed multi-area SE environment. 

The structure of this paper is as follows: Section II describes the conventional (centralized) SE and defines an optimization problem which allows a solution via the BP approach. Section III formulates closed form expressions for BP messages. In Section IV, we give a step-by-step description of the proposed algorithm, while Section V considers the convergence performance and numerical results for the IEEE 14 bus test case. Concluding remarks are included in Section VI. 

\section{Electric power system state estimation}
The AC SE problem reduces to solving the system of equations\cite{schweppe}:
		\begin{equation}
        \begin{aligned}
        \mathbf{z}=\mathbf{h}(\mathbf{x})+\mathbf{u},
        \end{aligned}
		\label{eqn1}
		\end{equation}
where $\mathbf{h}(\mathbf{x})=(h_1(\mathbf{x}), \dots, h_k(\mathbf{x}))$ includes both non-linear and linear measurement functions (see Appendix for details), $\mathbf {x}=(x_1,\dots,x_{n})$ is the vector of the state variables, $\mathbf{z} = (z_1,\dots,z_k)$ is the vector of independent measurements (where $n < k$), and $\mathbf{u} = (u_1,\dots,u_k)$ is the vector of measurement errors.

The state variables are bus voltage magnitudes and bus voltage angles, transformer magnitudes of turns ratio and transformer angles of turns ratio. Without loss of generality, in the rest of the paper, we observe bus voltage angles $\bm \uptheta=$ $(\theta_1,\dots,\theta_N)$ and bus voltage magnitudes $\mathbf V=$ $(V_1,\dots,V_N)$ as state variables $\mathbf x \equiv (\bm \uptheta,\mathbf V)$, where $N$ is the number of buses ($n=2N$).

Under the assumption that measurement errors $\mathbf u$ follow a zero-mean Gaussian distribution, the probability density function associated with the m-th measurement equals:
		\begin{equation}
        \begin{gathered}
        \mathcal{N}(z_m|\mathbf{x},\sigma_m^2) = \cfrac{1}{\sigma_m\sqrt{2\pi}} 
        \exp\Bigg\{\cfrac{[z_m-h_m(\mathbf{x})]^2}{2\sigma_m^2}\Bigg\},
        \end{gathered}
		\label{eqn2}
		\end{equation}
where $z_m$ is the value of the measurement, $\sigma_m^2$ is the measurement variance, and the function $h_m(\mathbf{x})$ connects the vector of state variables to the value of m-th measurement.

One can find the MAP solution to the SE problem via maximization of the likelihood function, which is defined via likelihoods of $k$ independent measurements:  
		\begin{equation}
        \begin{gathered}
		\hat{\mathbf x}=\mathrm{arg} \max_{\mathbf{x}}\mathcal{L}(\mathbf{z}|\mathbf{x})=\mathrm{arg} \max_{\mathbf{x}}  \prod_{h=1}^k \mathcal{N}(z_h|\mathbf{x},\sigma_h^2).
        \end{gathered}
		\label{eqn3}
		\end{equation}

The maximum likelihood estimator \eqref{eqn3} is equivalent to the weighted least-squares estimator whose solution can be found using the Gauss-Newton method:
		\begin{subequations}
		\renewcommand{\theequation}{\theparentequation.\arabic{equation}}
        \begin{gather}  
		\mathbf J (\mathbf x^\nu)^\mathrm{T} \mathbf W \mathbf J (\mathbf x^\nu)\Delta \mathbf x^\nu =
		\mathbf J (\mathbf x^\nu)^\mathrm{T}\mathbf W\mathbf r (\mathbf x^\nu)\label{eqn4.1}\\
		 \mathbf x^{\nu+1} = \mathbf x^\nu + \Delta \mathbf x^\nu, \label{eqn4.2}
        \end{gather}
		\end{subequations}
where $\nu$ is the iteration step, $\Delta \mathbf x^\nu\in \mathbb {R}^{n}$ is the vector of increments of the state variables, $\mathbf J (\mathbf x^\nu)\in \mathbb {R}^{k\mathrm{x}n}$ is the Jacobian matrix of measurement functions $\mathbf h (\mathbf x^\nu)$ (see Appendix for details), $\mathbf{W}\in \mathbb {R}^{k\mathrm{x}k}$ is a diagonal matrix containing inverses of measurement variances, and $\mathbf r (\mathbf x^\nu) = \mathbf{z} - \mathbf h (\mathbf x^\nu)$ is the vector of residuals\cite{monticelli}.

At each iteration $\nu$, the Gauss-Newton method returns a new estimate of $\mathbf{x}$, which in a given iteration may be observed as a constant vector. If the Jacobian matrix $\mathbf J (\mathbf x^\nu)$ has a full column rank, the equation \eqref{eqn4.1} represents the linear WLS solution of the minimization problem \cite{hansen}:   
		\begin{equation}
        \begin{gathered}
		\min_{\Delta \mathbf x^\nu} 
		||\mathbf W^{1/2}[\mathbf r (\mathbf x^\nu)-\mathbf J (\mathbf x^\nu)\Delta \mathbf x^\nu]||_2^2.
        \end{gathered}
		\label{eqn5a}
		\end{equation}
Hence, at each iteration $\nu$, we can consider system of linear equations:
		\begin{equation}
        \begin{aligned}
        \mathbf r (\mathbf x^\nu)=\mathbf{g}(\Delta \mathbf x^\nu)+\mathbf{u},
        \end{aligned}
		\label{eqn7}
		\end{equation}	
where $\mathbf{g}(\Delta \mathbf x^\nu)= \mathbf J (\mathbf x^\nu)\Delta \mathbf x^\nu$ comprises linear functions. The equation \eqref{eqn4.1} is the weighted normal equation for the minimization problem defined as \eqref{eqn5a}, or alternatively, equation \eqref{eqn4.1} is WLS solution of \eqref{eqn7}.

Consequently, the probability density function associated with the m-th measurement (i.e., the m-th residual component $r_m$) at any iteration $\nu$:
		\begin{equation}
		\begingroup\makeatletter\def\f@size{9}\check@mathfonts
        \begin{gathered}
        \mathcal{N}(r_m(\mathbf x^\nu)|{\Delta \mathbf x^\nu},\sigma_m^2) 
        = \cfrac{1}{\sigma_m\sqrt{2\pi}} 
        \exp\Bigg\{\cfrac{[r_m(\mathbf x^\nu)-g_m(\Delta \mathbf x^\nu)]^2}{2\sigma_m^2}\Bigg\}.
        \end{gathered}
        \endgroup
		\label{eqn8}
		\end{equation}

The MAP solution of \eqref{eqn3} can be redefined as an iterative optimization problem where, instead of solving \eqref{eqn4.1} and \eqref{eqn4.2}, we solve MAP (sub)problem:
		\begin{equation}
        \begin{aligned}
		\Delta \hat {\mathbf x}^\nu&=
		\mathrm{arg} \max_{\Delta\mathbf{x}^\nu}
		\mathcal{L}\Big(\mathbf{r}(\mathbf{x}^\nu)|\Delta\mathbf{x}^\nu\Big)\\ 
		&= \mathrm{arg} \max_{\Delta\mathbf{x}^\nu} 
		\prod_{h=1}^k \mathcal{N} \Big(r_h(\mathbf{x}^\nu)|\Delta\mathbf{x}^\nu,\sigma_h^2\Big)\\
		\mathbf{x}^{\nu+1} &= \mathbf{x}^\nu+ \Delta \hat {\mathbf x}^\nu.
        \end{aligned}
		\label{eqn9}
		\end{equation}
As we show next, the solution to the above MAP subproblem over increment variables $\Delta {\mathbf x}^\nu$ can be efficiently obtained using BP algorithm applied over the underlying factor graph. 

Note that, if the factor graph corresponding to the problem \eqref{eqn9} (see Section III) is a tree, the resulting BP algorithm provides a solution equal to the linear WLS solution $\Delta {\mathbf x}^\nu$ of \eqref{eqn4.1}. In general, if the factor graph contains loops, the BP solution of $\Delta \hat {\mathbf x}^\nu$ in each iteration $\nu$ (outer iteration loop) will be obtained via iterative BP algorithm (inner iteration loops). Every inner BP iteration loop $\rho=1,2,\dots,\tau(\nu)$ outputs $\Delta \hat {\mathbf x}^{\nu,\rho}$, where $\tau(\nu)$ is the number of inner BP iterations within outer iteration $\nu$. 
	
\section{Factor Graphs and BP algorithm} 
Factor graphs and BP algorithm are widely used tools for probabilistic inference \cite{pearl}, \cite{bishop}. In our scenario, FGs consist of variable nodes for every variable in the likelihood function and of factor nodes for each likelihood factor. Therefore, for the MAP subproblem defined in \eqref{eqn9}, the increments $\Delta \mathbf x$ of state variables $\mathbf x$ will appear as variable nodes, while residuals $\mathbf r$ of measurements $\mathbf z$ will define factor nodes. A factor node connects to a variable node if the increment variable is an argument of the corresponding function ${g_m}({\Delta \mathbf x})$, which is equivalent to say that the corresponding state variable is an argument of the measurement function $h_m(\mathbf x)$.

\begin{figure}[H]
\centering
\includegraphics[width=4.4cm]{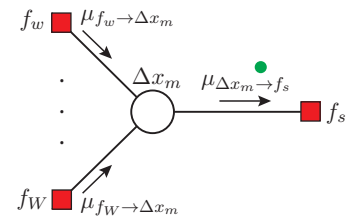}
\caption{Message from variable node $\Delta x_m$ to factor node $f_s$}
\label{fig1}
\end{figure}

Consider the part of a factor graph shown in Fig. \ref{fig1} with group of factor nodes $\mathcal{F}=\{ f_s,f_w,...,f_W\}$ that are neighbours of the variable node $\Delta x_m$. 
 \noindent
The message from the variable node $\Delta x_m$ to the factor node $f_{s}$ is equal to the product of all incoming factor node to variable node messages arriving at all the other incident edges: 
		\begin{equation}
        \begin{gathered}
        \mu_{\Delta x_m \to f_{s}}(\Delta x_m) =\prod_{f_a \in \mathcal{F} \setminus f_{s}} 
        \mu_{f_{a} \to \Delta x_m}(\Delta x_m),
        \end{gathered}
		\label{eqn12}
		\end{equation}
where $\mathcal{F} \setminus f_{s}$ defines the set of factor nodes incident to the variable node $\Delta x_m$, excluding the factor node $f_{s}$. 		

It can be shown that the message $\mu_{\Delta x_m \to f_{s}}(\Delta x_m)$ is represented by the Gaussian function: 
		\begin{equation}
        \begin{gathered}
        \mu_{\Delta x_m \to f_{s}}(\Delta x_m) \propto
        \mathcal{N}(r_{\Delta x_m \to f_{s}}|\Delta x_m,\sigma_{\Delta x_m \to f_{s}}^2),
        \end{gathered}
		\label{eqn13}
		\end{equation}	
with mean $r_{\Delta x_m \to f_{s}}$ and variance $\sigma_{\Delta x_m \to f_{s}}^2$:
		\begin{equation}
        \begin{aligned}
        r_{\Delta x_m \to f_{s}} &= 
        \Bigg( \sum_{f_a \in \mathcal{F}\setminus f_{s}} \cfrac{r_{f_{a} \to \Delta x_m}}
        {\sigma_{f_{a} \to \Delta x_m}^2}\Bigg)
        \sigma_{\Delta x_m \to f_{s}}^2\\
        \cfrac{1}{\sigma_{\Delta x_m \to f_{s}}^2} &= 
		\sum_{f_a \in \mathcal{F}\setminus f_{s}} \cfrac{1}{\sigma_{f_{a} \to \Delta x_m}^2}. 
        \end{aligned}
		\label{eqn13a}
		\end{equation}	

\begin{figure}[H]
\centering
\includegraphics[width=5.0cm]{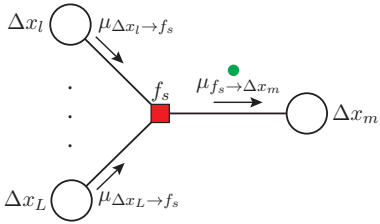}
\caption{Message from factor node $f_s$ to variable node $\Delta x_m$}
\label{fig2}
\end{figure}

Consider the part of the factor graph shown in Fig. \ref{fig2} that consists of the group of variable nodes $\mathcal{X} \in \{ \Delta x_m, \Delta x_l,...,\Delta x_L\}$ that are neighbours of the factor node $f_s$. The message from the factor node $f_s$ to the variable node $\Delta x_m$ is defined as a product of all incoming variable node to factor node messages arriving at all the other incident edges multiplied by the Gaussian function associated to the factor node $f_s$ and marginalized over all of the variables associated with the incoming messages:
		\begin{equation}
		\begingroup\makeatletter\def\f@size{9}\check@mathfonts
        \begin{aligned}
        \mu_{f_s \to \Delta x_m}(\Delta x_m)= 
		\int\displaylimits_{\Delta x_l}\dots\int\displaylimits_{\Delta x_L} 
		\mathcal{N}(r_{f_s}|\Delta x_m,\Delta x_l\dots \Delta x_L,\sigma_{f_s}^2)\\
		\prod_{\Delta x_b \in \mathcal{X}\setminus \Delta x_m} \mu_{\Delta x_b \to f_s}(\Delta x_b) \cdot \mathrm{d}\Delta x_b, 
        \end{aligned}
		\endgroup
		\label{eqn14}
		\end{equation}		
where $\mathcal{X}\setminus \Delta x_m$ is the set of variable nodes incident to the factor node $f_s$, excluding the variable node $\Delta x_m$.

 \noindent
It can be shown that the message $ \mu_{f_s \to \Delta x_m}(\Delta x_m)$ is represented by the Gaussian function:
		\begin{equation}
        \begin{gathered}
        \mu_{f_s \to \Delta x_m}(\Delta x_m) \propto
        \mathcal{N}(r_{f_s \to \Delta x_m}|\Delta x_m,\sigma_{f_s \to \Delta x_m}^2),
        \end{gathered}
		\label{eqn15}
		\end{equation}
with mean $r_{f_s \to \Delta x_m}$ and variance $\sigma_{f_s \to \Delta x_m}^2$:		
		\begin{equation}
        \begin{gathered}
		r_{f_s \to \Delta x_m}=
		\cfrac{1}{C_{\Delta x_m}}
		\Bigg( r_{f_s} - \sum_{\Delta x_b \in \mathcal{X} \setminus \Delta x_m} 
		C_{\Delta x_b} \cdot r_{\Delta x_b \to f_s}  
		 \Bigg)\\		
		\sigma_{f_s \to \Delta x_m}^2 = 
		\cfrac{1}{C_{\Delta x_m}^2}
		\Bigg( \sigma_{f_s}^2 + \sum_{\Delta x_b \in \mathcal{X} \setminus \Delta x_m} 
		C_{\Delta x_b}^2 \cdot \sigma_{\Delta x_b \to f_s}^2  
		 \Bigg).
        \end{gathered}
		\label{eqn15a}
		\end{equation}
The coefficients $C_{\Delta x_i},\; i=m,l\dots,L$, are Jacobian elements of the measurement function (see Appendix for details) associated with the factor node $f_s$: 
		\begin{equation}
        \begin{gathered}
		C_{\Delta x_i}=\cfrac{\partial h(x_m,x_l,\dots, x_L)}{\partial x_i},\;i=m,l\dots,L.	
        \end{gathered}
		\label{eqn15j}
		\end{equation}

Note that, due to the fact that all the measurements follow Gaussian distribution and that BP processing in both variable and function nodes preserve "Gaussianity", all the messages exchanged in the presented BP are Gaussian distributions. The resulting BP algorithm is known as Gaussian BP algorithm in which all the BP messages are completely represented using only means and variances\cite{ping}. 

\begin{figure}[H]
\centering
\includegraphics[width=4.5cm]{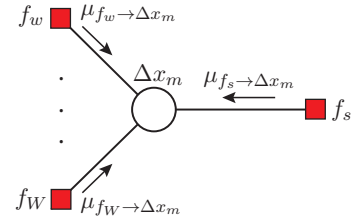}
\caption{Marginal inference for $\Delta x_m$}
\label{fig3}
\end{figure}

The marginal of state variable increment $\Delta x_m$, illustrated in Fig. \ref{fig3}, is obtained as the product of all incoming messages into the variable node $\Delta x_m$:
		\begin{equation}
        \begin{gathered}
        p(\Delta x_m) =\prod_{f_c \in \mathcal{F}} \mu_{f_{c} \to \Delta x_m}(\Delta x_m),
        \end{gathered}
		\label{eqn16}
		\end{equation}
where $\mathcal{F}$ is the set of factor nodes incident to the variable node $\Delta x_m$.		
 \noindent
Thus the marginal has Gaussian form: 
		\begin{equation}
        \begin{gathered}
        p(\Delta x_m) \propto 
        \mathcal{N}(\Delta \hat x_m|\Delta x_m,\sigma_{\Delta x_m}^2), 
        \end{gathered}
		\label{eqn17}
		\end{equation}
with mean $\Delta \hat x_m$ which represents the estimated value of the state variable increment $\Delta x_m$ and variance $\sigma_{\Delta x_m}^2$:
		\begin{equation}
        \begin{aligned}
        \Delta \hat x_m &= 
        \Bigg( \sum_{f_c \in \mathcal{F}} \cfrac{r_{f_{c} \to \Delta x_m}}
        {\sigma_{f_{c} \to \Delta x_m}^2}\Bigg)
        \sigma_{\Delta x_m}^2\\
        \cfrac{1}{\sigma_{\Delta x_m}^2} &= 
		\sum_{f_c \in \mathcal{F}} \cfrac{1}{\sigma_{f_{c} \to \Delta x_m}^2}.  
        \end{aligned}
		\label{eqn17a}
		\end{equation}

The MAP subproblem defined in \eqref{eqn9} can be efficiently solved using \eqref{eqn13a}, \eqref{eqn15a} and \eqref{eqn17a}. 

\section{Toy Example}
\begin{figure}[H]
\centering
\includegraphics[width=50mm]{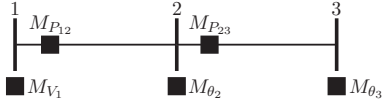}
\caption{Bus/branch model}
\label{fig4}
\end{figure}

An illustrative example presented in Fig. \ref{fig4} will be used to provide a step-by-step presentation of the proposed algorithm. The simple three bus radial network contains three direct measurement devices that directly measure state variables $M_{dir} \in \{ M_{V_1}, M_{\theta_2},M_{\theta_3} \}$, and two indirect measurement devices $M_{ind} \in \{M_{P_{12}},M_{P_{23}} \}$ that measure state variables indirectly. 
 \noindent

Input data for SE from measurement devices are Gaussian-type functions represented by means and variances: $z_{dir} \in \{ z_{V_1}, z_{\theta_2},z_{\theta_3} \}$, $\sigma_{dir}^2 \in \{ \sigma_{V_1}^2, \sigma_{\theta_2}^2,\sigma_{\theta_3}^2 \}$ and $z_{ind} \in \{ z_{P_{12}}, z_{P_{23}} \}$ , $\sigma_{ind}^2 \in \{ \sigma_{P_{12}}^2, \sigma_{P_{23}}^2 \}$.

\begin{figure}[H]
\centering
\includegraphics[width=75mm]{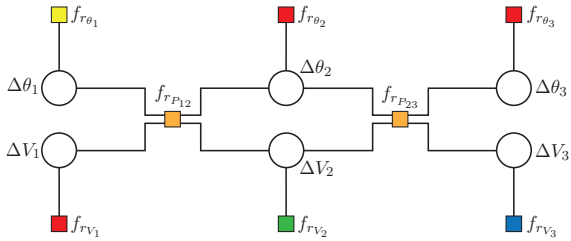}
\caption{Factor graph of the illustrative example}
\label{fig5}
\end{figure}

The corresponding FG is given in Fig. \ref{fig5}, where we define indirect factor nodes $f_{r_{P_{12}}}$,  $f_{r_{P_{23}}}$ (orange squares) corresponding to indirect measurements and four types of singly-connected factor nodes (local factor nodes) described next. The slack factor node $f_{r_{\theta_1}}$ (yellow square) corresponds to the slack or reference bus where the voltage angle has a given value, therefore, the residual of that state variable is equal to zero. The direct factor nodes $f_{r_{V_1}}$, $f_{r_{\theta_2}}$, $f_{r_{\theta_3}}$ (red squares) correspond to the direct measurements. The initialization factor node $f_{r_{V_2}}$ (green square) is needed to start the algorithm, while the virtual factor node $f_{r_{V_3}}$ (blue square) is used to form a message from a variable node to a factor node. In general, if the variable node is not directly measured and is singly-connected to the rest of the FG, we attach a virtual factor node to this variable node. 
	
 \noindent

In the following, for notational convenience, we denote the variance as follows: $\sigma^2 \equiv \xi$.	

\noindent
\textbf{Algorithm Initialization}
\begin{enumerate}
\item The AC SE in electric power systems assumes "flat start" or a priori given values of state variables: 
	 	\begin{equation}
        \begin{gathered}
        \mathbf x^{\{ \nu=0 \} }=[\theta_1\; \theta_2\; \theta_3\; V_1\; V_2\; V_3]^{\{ \nu=0 \} }.
        \end{gathered}
		\label{toy1} \nonumber
		\end{equation}
\item The value of the slack factor node is set to $r_{\theta_1}=0$ with variance $\xi_{\theta_1} \to 0$. 
\item The value of initialization factor nodes and virtual factor nodes are set to $r_{V_2} \to 0$ and $r_{V_3} \to 0$, with variances $\xi_{V_2} \to \infty$ and $\xi_{V_3} \to \infty$.
\end{enumerate}

\noindent
\textbf{Iterate - Outer Loop:} $\bm\nu \mathbf{=0,1,2,\dots}; \bm\rho \mathbf{=0 }$
\begin{enumerate}[resume]			
\item Each direct factor node computes residual, e.g.:
	 	\begin{equation}
        \begin{gathered}
		r_{\theta_2}^{\{ \nu \} } = z_{\theta_2}-\theta_2^{\{ \nu \} }.        
		\end{gathered} 
		\label{toy2} \nonumber
		\end{equation}	

\item Local factor nodes send messages represented by a triplet (residual, variance, state variable), to incident variable nodes, e.g.:
	 	\begin{equation}
        \begin{gathered}
		\mu_{f_{r_{\theta_1}} \to \Delta \theta_1}^{\{ \nu \} } :=
		\big(r_{\theta_1},\xi_{\theta_1},\theta_1^{\{ \nu \} }\big)\\
		\mu_{f_{r_{\theta_2}} \to \Delta \theta_2}^{\{ \nu \} } := 
		\big(r_{\theta_2}^{\{ \nu \} },\xi_{\theta_2},\theta_2^{\{ \nu \} }\big).        
		\end{gathered}
		\label{toy3} \nonumber
		\end{equation}
\item Variable nodes forward the incoming messages received from local factor nodes along remaining edges, e.g.: 
	 	\begin{equation}
        \begin{aligned}
		\mu_{\Delta \theta_2 \to f_{r_{P_{12}}}}^{\{ \nu \}\{ \rho \} } &= 
		\mu_{f_{r_{\theta_2}} \to \Delta \theta_2}^{\{ \nu \} }:= 
		\big(r_{\theta_2}^{\{ \nu \} },\xi_{\theta_2},\theta_2^{\{ \nu \} }\big)\\
		&:=\big(r_{\Delta \theta_2 \to f_{r_{P_{12}}}}^{\{ \nu \}\{ \rho \} },
		\xi_{\Delta \theta_2 \to f_{r_{P_{12}}}}^{\{ \nu \}\{ \rho \} },
		\theta_2^{\{ \nu \}\{ \rho \} }\big) \\[2ex]
		\mu_{\Delta \theta_2 \to f_{r_{P_{23}}}}^{\{ \nu \}\{ \rho \} } &= 
		\mu_{f_{r_{\theta_2}} \to \Delta \theta_2}^{\{ \nu \} }:= 
		\big(r_{\theta_2}^{\{ \nu \} },\xi_{\theta_2},\theta_2^{\{ \nu \} }\big)\\
		&:=\big(r_{\Delta \theta_2 \to f_{r_{P_{23}}}}^{\{ \nu \}\{ \rho \} },
		\xi_{\Delta \theta_2 \to f_{r_{P_{23}}}}^{\{ \nu \}\{ \rho \} },
		\theta_2^{\{ \nu \}\{ \rho \} }\big). 	
		\end{aligned}
		\label{toy4} \nonumber
		\end{equation}	
\item Indirect factor nodes compute residuals, e.g.:
	 	\begin{equation}
        \begin{gathered}
		r_{P_{12}}^{\{ \nu \} }=z_{P_{12}}-
		h_{P_{12}}(\theta_1^{\{ \nu \} }, \theta_2^{\{ \nu \} }, V_1^{\{ \nu \} }, V_2^{\{ \nu \} }).			
		\end{gathered}
		\label{toy5} \nonumber
		\end{equation}
\item Indirect factor nodes compute appropriate Jacobian elements associated with state variables, e.g.: 
		\begin{equation}
        \begingroup\makeatletter\def\f@size{8.8}\check@mathfonts
        \begin{aligned}
     	C_{P_{12},\Delta \theta_1}^{\{ \nu \} }=
     	\cfrac{\mathrm \partial{h_{P_{12}}(\cdot)}}{\mathrm \partial \theta_{1}}&=
     	{V}_{1}^{\{ \nu \} }{V}_{2}^{\{ \nu \} }
     	(g_{12}\mbox{sin}\theta_{12}^{\{ \nu \} }-b_{12}\mbox{cos}\theta_{12}^{\{ \nu \} })\\
     	C_{P_{12},\Delta V_2}^{\{ \nu \} }=
     	\cfrac{\mathrm \partial{h_{{P_{12}}}(\cdot)}}{\mathrm \partial V_{2}}&=
     	-{V}_{1}^{\{ \nu \} }(g_{12}\mbox{cos}\theta_{12}^{\{ \nu \} }+b_{12}\mbox{sin}\theta_{12}^{\{ \nu \} }).
        \end{aligned}
     	\endgroup        
		\label{toy6} \nonumber
		\end{equation}	
\end{enumerate}

\noindent
\textbf{Iterate - Inner Loop:} $\bm\rho \mathbf{=1,2,\dots, \bm\tau(\bm\nu)}$
\begin{enumerate}[resume]	
\item Indirect factor nodes send messages as pairs along incident edges according to \eqref{eqn15a}, e.g.:
  	 	\begin{equation}
        \begin{aligned}
		\mu_{f_{r_{P_{12}}} \to \Delta \theta_2}^{\{ \rho \} } := 
		\big(r_{f_{r_{P_{12}}} \to \Delta \theta_2}^{\{ \rho \} } ,
		\xi_{f_{r_{P_{12}}} \to \Delta \theta_2}^{\{ \rho \} }\big)
		\end{aligned}
		\label{} \nonumber
		\end{equation}
		\begin{equation}
        \begin{aligned}
		r_{f_{r_{P_{12}}} \to \Delta \theta_2}^{ \{ \rho \} } =
		\cfrac{1}{C_{P_{12},\Delta \theta_2}^{ \{ \nu \} }}\Big[
		r_{P_{12}}^{ \{ \nu \} }-
		C_{P_{12},\Delta \theta_1}^{ \{ \nu \} }\cdot
		r_{\Delta \theta_1 \to f_{r_{P_{12}}}}^{\{ \nu \}\{ \rho-1 \} }\\
		-C_{P_{12},\Delta V_1}^{ \{ \nu \} }\cdot
		r_{\Delta V_1 \to f_{r_{P_{12}}}}^{\{ \nu \}\{ \rho-1 \} }
		C_{P_{12},\Delta V_2}^{ \{ \nu \} }\cdot
		r_{\Delta V_2 \to f_{r_{P_{12}}}}^{\{ \nu \}\{ \rho-1 \} }
		\Big] 
		\end{aligned}
		\label{toy7} \nonumber
		\end{equation}\\[-4ex]
		\begin{equation}
        \begin{aligned}
		\xi_{f_{r_{P_{12}}} \to \Delta \theta_2}^{\{ \rho \} } =
		\cfrac{1}{(C_{P_{12},\Delta \theta_2}^{\{ \nu \} })^2}\Big[
		\xi_{P_{12}}+
		(C_{P_{12},\Delta \theta_1}^{ \{ \nu \} })^2\cdot
		\xi_{\Delta \theta_1 \to f_{r_{P_{12}}}}^{\{ \nu \}\{ \rho-1 \} }\\
		+(C_{P_{12},\Delta V_1}^{ \{ \nu \} })^2\cdot
		\xi_{\Delta V_1 \to f_{r_{P_{12}}}}^{\{ \nu \}\{ \rho-1 \} }+
		({C_{P_{12},\Delta V_2}^{\{ \nu \} }})^2\cdot
		\xi_{\Delta V_2 \to f_{r_{P_{12}}}}^{\{ \nu \}\{ \rho-1 \} }
		\Big].	
		\end{aligned}
		\label{toy7} \nonumber
		\end{equation}
		
\pagebreak		
\item Variable nodes send messages as pairs along incident edges to indirect factor nodes according to \eqref{eqn13a}, e.g.:
  	 	\begin{equation}
        \begin{gathered}
		\mu_{\Delta \theta_2 \to f_{r_{P_{12}}}}^{\{ \nu \}\{ \rho \} } :=
		\big(r_{\Delta \theta_2 \to f_{r_{P_{12}}}}^{\{ \nu \}\{ \rho \} } ,
		\xi_{\Delta \theta_2 \to f_{r_{P_{12}}}}^{\{ \nu \} \{ \rho \} }\big)
		\end{gathered}
		\label{toy8} \nonumber
		\end{equation}
		\begin{equation}
        \begin{aligned} 
		\cfrac{1}{\xi_{\Delta \theta_2 \to f_{r_{P_{12}}}}^{\{ \nu \}\{ \rho \} }}&=
		\cfrac{1}{\xi_{\theta_2}}+
		\cfrac{1}{\xi_{f_{r_{P_{23}}} \to \Delta \theta_2}^{ \{ \rho \} }}\\
		r_{\Delta \theta_2 \to f_{r_{P_{12}}}}^{\{ \nu \}\{ \rho \} }&=\Bigg(
		\cfrac{r_{\theta_2}^{ \{ \nu \}}}{\xi_{\theta_2}}+
		\cfrac{r_{f_{r_{P_{23}}} \to \Delta \theta_2}^{ \{ \rho \} }}
		{\xi_{f_{r_{P_{23}}} \to \Delta \theta_2}^{ \{ \rho \} }} \Bigg)
		\xi_{\Delta \theta_2 \to f_{r_{P_{12}}}}^{ \{ \rho \} }.
		\end{aligned}
		\label{toy8} \nonumber
		\end{equation}	
\end{enumerate}

\noindent
\textbf{Iterate - Outer Loop:} $\bm\nu \mathbf{=0,1,2,\dots}; \bm\rho =\bm{\tau(\bm\nu) }$
\begin{enumerate}[resume]	
\item Variable nodes compute marginals according to \eqref{eqn17a}, e.g.:
  	 	\begin{equation}
        \begin{gathered}
        p(\Delta \theta_2) \propto 
        \mathcal{N}(\Delta \hat \theta_2|\Delta \theta_2, \hat \xi_{\theta_2})
		\end{gathered}
		\label{toy9} \nonumber
		\end{equation}	
  	 	\begin{equation}
        \begin{aligned}
        \cfrac{1}{\hat \xi_{\Delta \theta_2}^{\{ \nu \} }}&=
		\cfrac{1}{\xi_{\theta_2}}+
		\cfrac{1}{\xi_{f_{r_{P_{12}}} \to \Delta \theta_2}^{ \{ \rho \} }}+
		\cfrac{1}{\xi_{f_{r_{P_{23}}} \to \Delta \theta_2}^{ \{ \rho \} }}\\
		\Delta \hat \theta_2^{\{ \nu \} }&=\Bigg(
		\cfrac{r_{\theta_2}^{ \{ \nu \} }}{\xi_{\theta_2}}+
		\cfrac{r_{f_{r_{P_{12}}} \to \Delta \theta_2}^{ \{ \rho \} }}
		{\xi_{f_{r_{P_{12}}} \to \Delta \theta_2}^{ \{ \rho \} }}+
		\cfrac{r_{f_{r_{P_{23}}} \to \Delta \theta_2}^{ \{ \rho \} }}
		{\xi_{f_{r_{P_{23}}} \to \Delta \theta_2}^{ \{ \rho \} }} \Bigg)
		\hat \xi_{\Delta \theta_2}^{ \{ \nu \} }.
		\end{aligned}
		\label{toy9} \nonumber
		\end{equation}	 		
\item Variable nodes update the state variables, e.g.:
  	 	\begin{equation}
        \begin{gathered}
        \theta_2^{ \{ \nu+1 \} } = \theta_2^{ \{ \nu \} }+\Delta \hat \theta_2^{ \{ \nu \} }.
		\end{gathered}
		\label{toy10} \nonumber
		\end{equation}
\item Repeat steps 4-13 until convergence\footnote{Note that, after step 8,, initialization factor nodes are removed from the FG. Also in each iteration, virtual factor nodes repeat the same message as in the initial step and messages from a virtual factor node to a variable node should not be included in calculation of the marginals.}. 		
\end{enumerate}

\section{Numerical Results}
\begin{figure}[H]
	\centering
	\includegraphics[width=75mm]{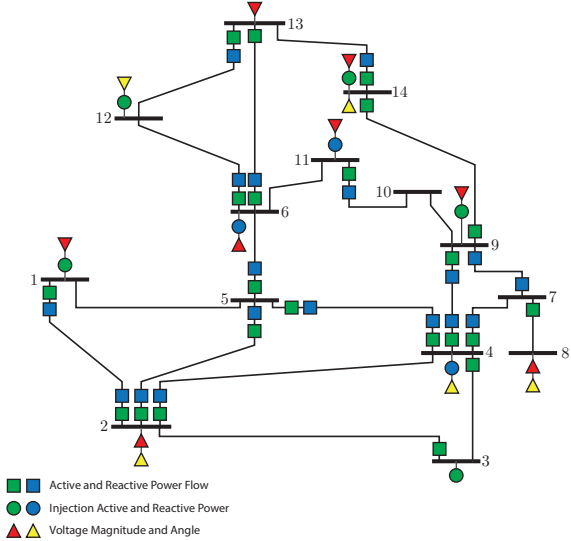}
	\caption{The IEEE 14 bus test case}
	\label{fig6}
	\end{figure}

The IEEE 14 bus test case shown in Fig. \ref{fig6} is used to analyse performance of the proposed algorithm. The set of measurements obtained from 61 measurement devices that measure active and reactive power flow, active and reactive injection power, bus voltage magnitude and bus voltage angle are selected so that the system is observable.
	 \noindent

\subsection{Simulation setup}
From a given IEEE 14 bus test case and the set of measurements, we define the corresponding factor graph. Measurement values are generated using the AC power flow analysis, additionally corrupted by Gaussian white noise of variance $\sigma^2$. For each value of variance $\sigma^2$, using Monte Carlo approach, we generate 1000 random sets of measurement values and feed them to the proposed BP-based SE algorithm in order to obtain the average performance results. 

The convergence of the proposed algorithm is tracked by observing the root mean square error (RMSE) after each outer iteration:
	\begin{equation}
	\begin{gathered}
	\mathrm{RMSE}(\nu) = { \cfrac{1}{n} ||\mathbf {\hat x}- \mathbf{x}^{\nu}||_2 },
	\end{gathered}
	\label{eqn22}
	\end{equation}
where $\mathbf {\hat x}$ is the non-linear WLS solution, while $\mathbf{x}^{\nu}$ represents the current iterate solution of the BP algorithm.

We investigate two simulation scenarios using "flat start" ($V_i=1$, $\theta_i=0$, $i=1,\dots,N$). In the first scenario, we test the algorithm convergence by measuring RMSE using the described Monte Carlo approach for different values of measurement variances from the set $\sigma^2=$ $\{\sigma_1^2$, $\sigma_2^2$, $\sigma_3^2$, $\sigma_4^2 \}$ $=\{0.01^2$, $0.001^2$, $0.0001^2$, $0.00001^2 \} \; [\mbox{p.u.}]$. For every outer iteration $\nu$, the number of inner iterations is defined as $\tau(\nu) = \nu^q$, where $q$ is inner iteration exponent, which we set to $q=4$. In the second scenario, we change the inner iteration number exponent $q$ while keeping the variance fixed at the values $\sigma_1^2$ (high noise level) and $\sigma_4^2$ (low noise level). The convergence is discussed and presented in the following subsection.

\subsection{Simulation results}
The set of measurements defines the topology of the FG and for almost all placements of measurement devices of interest, the corresponding FG will have loops\footnote{Note that, even if the physical power network has the radial structure (tree structure), the FG will be loopy. An exception occurs, for example, for the scenario of a radial network in which only end buses (and no internal buses on a radial line) are allowed to contain power injection measurements.}. It is well known that, in general, loopy BP does not converge to correct marginals, e.g., specific inputs may lead to an oscillatory behaviour of messages \cite{jordan}. Based on extensive numerical studies, we presented in \cite{extended} a heuristic solution to improve the convergence of the BP algorithm:
		\begin{equation}
        \begin{gathered}
		\mu_{f \to x}^\rho=
		[1-\delta(p)]\cdot\mu_{f \to x}^\rho + \delta(p)\cdot 
		\alpha \cdot [\mu_{f \to x}^{\rho-1}+\mu_{f \to x}^\rho], 
		\end{gathered}
		\label{eqn22a}
		\end{equation}
where $\delta(p) \in \{0,1\}$ is a Bernoulli random variable with parameter $p$, independently sampled for each message $\mu_{f \to x}^\rho$, and $\alpha$ is the weighting coefficient. Namely, we modify updates of factor to variable node messages in the inner iteration loop $\mu_{f \to x}^\rho$. For the values of $p\in [0.4,0.6]$ and $\alpha=0.5$, our numerical studies showed that the BP algorithm always converged successfully to the WLS solution. 

\begin{figure}[H]
	\centering
	\includegraphics[width=65mm]{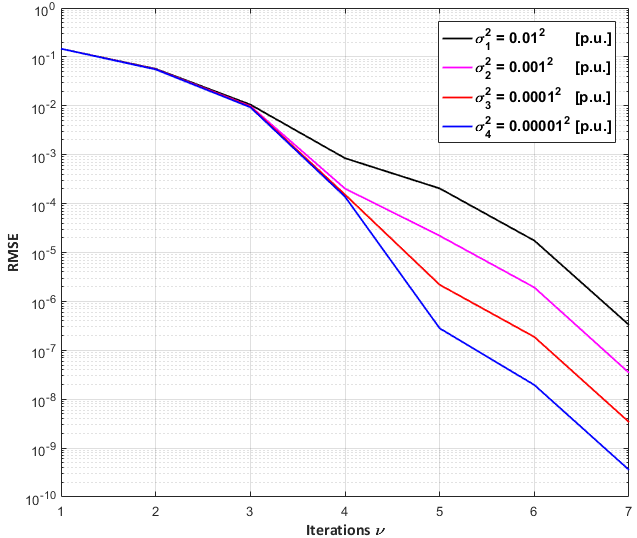}
	\caption{Convergence performance for different variances $\sigma^2$}
	\label{fig7}
	\end{figure}

\begin{figure}[H]
\centering
\includegraphics[width=65mm]{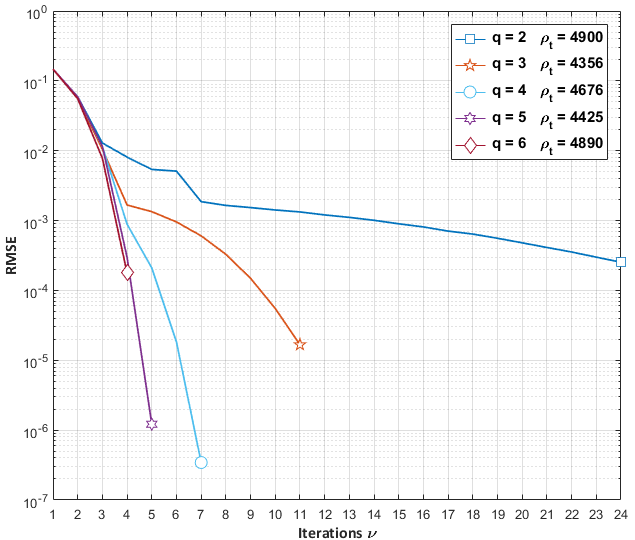}
\caption{Convergence performance for different $q$ and $\sigma_1^2$}
\label{fig8}
\end{figure}

\begin{figure}[H]
\centering
\includegraphics[width=65mm]{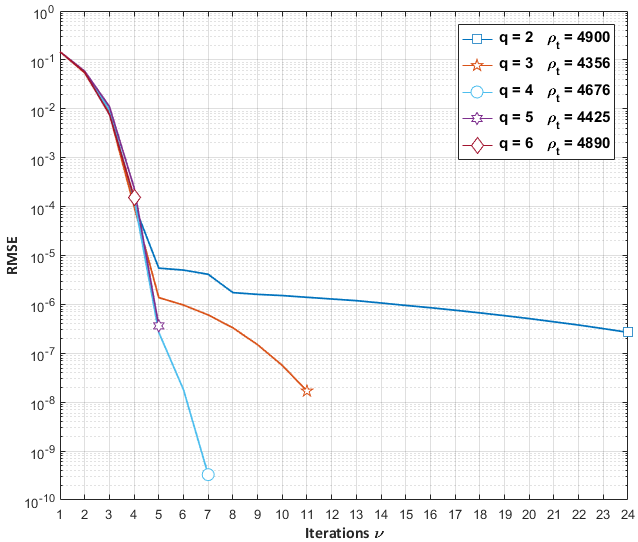}
\caption{Convergence performance for different $q$ and $\sigma_4^2$}
\label{fig9}
\end{figure}

Fig. \ref{fig7} shows the convergence behaviour for the first simulation scenario described in the previous subsection. The algorithm is terminated after $\nu = 7$ outer iterations with the total number of inner iterations $\rho_t =\sum_{\nu=1}^7 \nu^4= 4676$. The figure demonstrates that the proposed algorithm converges to WLS solution over a wide range of noise variances. As expected, the convergence behaviour improves as the noise levels decreases. 
	 \noindent
The convergence behaviour for different values of $q$ for the set of measurements with variances $\sigma_1^2$ (high noise level) and $\sigma_4^2$ (low noise level) is shown in Fig. \ref{fig8} and Fig. \ref{fig9}.  

 \noindent
For different values of $q$, the number of outer iterations $\nu$ is selected in such a way that the resulting number of inner iterations $\rho$ is approximately the same. For both high and low noise levels, the value of inner iteration number exponent $q=4$ is identified to provide fastest convergence to the WLS solution for a given number of inner iterations. Note that for insufficient value of exponent $q$ (e.g., $q=2$), the convergence speed of the proposed algorithm will be dramatically reduced. The fastest convergence behaviour is obtained for $q=4$ (although $q=5$ also shows good performance). For too large $q$ value ($q \geq 6$), the convergence speed will decrease as compared to $q=4$.
  \noindent

\section{Conclusion}
In this paper, we presented the BP solution of the AC SE problem that can be interpreted as a fully distributed Gauss-Newton method. The proposed BP algorithm converges to the same solution as the centralized WLS state estimator. For the future work, we plan to compare the proposed algorithm in the multi-area SE setup in terms of performance, convergence and computational cost with the solutions recently proposed in the literature.

\section*{Acknowledgment}
This project has received funding from the EU 7th Framework Programme for research, technological development and demonstration under grant agreement no. 607774.

\section*{APPENDIX\\Measurement functions and Jacobian elements}
The measurement functions $\mathbf{h}(\mathbf{x}) \equiv \mathbf{h}(\mathbf V, \bm \uptheta)$ that connect measurements $\mathbf z$ to state variables $\mathbf x \equiv (\bm \uptheta,\mathbf V)$ and corresponding Jacobian elements are described below.

The measurement function for \textbf{active power flow} at the branch that connects buses $i$ and $j$: 
		\begin{equation}
        \begin{aligned}
        h_{P_{ij}}(\cdot) &= V_i^2(g_{ij}+g_{si})-V_iV_j(g_{ij}\cos\theta_{ij}+b_{ij}\sin\theta_{ij}),
        \end{aligned}
		\label{Ae1}\nonumber
		\end{equation}
where $V_i$ and $V_j$ are bus voltage magnitudes, while $\theta_{ij} = \theta_i - \theta_j$ is the bus voltage angle difference between buses $i$ and $j$. The parameters in above equations include the conductance $g_{ij}$ and susceptance $b_{ij}$ of the branch, as well as the conductance $g_{si}$ of the branch shunt element connected at the bus $i$. The Jacobian expressions corresponding to $h_{P_{ij}}(\cdot)$ are as follows:	
		\begin{align*}
     	\cfrac{\mathrm \partial{h_{P_{ij}}(\cdot)}}{\mathrm \partial \theta_{i}}&=
     	{V}_{i}{V}_{j}(g_{ij}\mbox{sin}\theta_{ij}-b_{ij}\mbox{cos}\theta_{ij})\\
     	\cfrac{\mathrm \partial{{h_{P_{ij}}}(\cdot)}}{\mathrm \partial \theta_{j}}&=
     	-{V}_{i}{V}_{j}(g_{ij}\mbox{sin}\theta_{ij}-b_{ij}\mbox{cos}\theta_{ij})\\
     	\cfrac{\mathrm \partial{{h_{P_{ij}}}(\cdot)}}{\mathrm \partial V_{i}}&=
    	-{V}_{j}(g_{ij}\mbox{cos}\theta_{ij}+b_{ij}\mbox{sin}\theta_{ij})+2V_{i}(g_{ij}+g_{si})\\
     	\cfrac{\mathrm \partial{h_{{P_{ij}}}(\cdot)}}{\mathrm \partial V_{j}}&=
     	-{V}_{i}(g_{ij}\mbox{cos}\theta_{ij}+b_{ij}\mbox{sin}\theta_{ij}).
        \end{align*}

The measurement function for \textbf{reactive power flow} at the branch that connects buses $i$ and $j$: 
		\begin{equation}
        \begin{aligned}
        h_{Q_{ij}}(\cdot) &= -V_i^2(b_{ij}+b_{si})-V_iV_j(g_{ij}\sin\theta_{ij}-b_{ij}\cos\theta_{ij}),
        \end{aligned}
		\label{Ae1-1}\nonumber
		\end{equation}
where $b_{si}$ is susceptance of the branch shunt element connected at the bus $i$. The Jacobian expressions corresponding to $h_{Q_{ij}}(\cdot)$ are as follows:		
		\begin{align*}
     	\cfrac{\mathrm \partial{h_{{Q_{ij}}}(\cdot)}}{\mathrm \partial \theta_{i}}&=
     	-{V}_{i}{V}_{j}(g_{ij}\mbox{cos}\theta_{ij}+b_{ij}\mbox{sin}\theta_{ij})\\
     	\cfrac{\mathrm \partial{h_{{Q_{ij}}}(\cdot)}}{\mathrm \partial \theta_{j}}&=
     	{V}_{i}{V}_{j}(g_{ij}\mbox{cos}\theta_{ij}+b_{ij}\mbox{sin}\theta_{ij})\\
     	\cfrac{\mathrm \partial{h_{{Q_{ij}}}(\cdot)}}{\mathrm \partial V_{i}}&=
     	-{V}_{j}(g_{ij}\mbox{sin}\theta_{ij}-b_{ij}\mbox{cos}\theta_{ij})-2V_{i}(b_{ij}+b_{si})\\
     	\cfrac{\mathrm \partial{h_{{Q_{ij}}}(\cdot)}}{\mathrm \partial V_{j}}&=-
     	{V}_{i}(g_{ij}\mbox{sin}\theta_{ij}-b_{ij}\mbox{cos}\theta_{ij}).
        \end{align*}

The measurement function for \textbf{active injection power} into the bus $i$:	
		\begin{equation}
        \begin{aligned}
        h_{P_{i}}(\cdot) &= 
        V_i \sum_{j \in \mathcal{H}} V_j(G_{ij}\cos\theta_{ij}+B_{ij}\sin\theta_{ij}),
        \end{aligned}
		\label{Ae2}\nonumber
		\end{equation} 
where $\mathcal{H}$ is the set of buses incident to the bus $i$, including the bus $i$. The parameters $G_{ij}$ and $B_{ij}$ are conductance and susceptance of the complex bus matrix. The Jacobian expressions corresponding to $h_{P_{i}}(\cdot)$ are:
\begin{equation}
        \begin{aligned}
     	\cfrac{\mathrm \partial{h_{P_{i}}(\cdot)}}{\mathrm \partial \theta_{i}}&=
     	{V}_{i}\sum_{j \in \mathcal{H} \setminus i } {V}_{j}
     	(-G_{ij}\mbox{sin}\theta_{ij}+B_{ij}\mbox{cos}\theta_{ij})\\
     	 \cfrac{\mathrm \partial{h_{P_{i}}(\cdot)}}{\mathrm \partial \theta_{j}}&=
     	{V}_{i}{V}_{j}(G_{ij}\mbox{sin}\theta_{ij}-B_{ij}\mbox{cos}\theta_{ij})\\
     	\cfrac{\mathrm \partial{h_{P_{i}}(\cdot)}}{\mathrm \partial V_{i}}&=
     	\sum_{j \in \mathcal{H} \setminus i } 
     	{V}_{j}(G_{ij}\mbox{cos}\theta_{ij}+B_{ij}\mbox{sin}\theta_{ij})+2{V}_{i}G_{ii}\\
     	\cfrac{\mathrm \partial{h_{P_{i}}(\cdot)}}{\mathrm \partial V_{j}}&=
     	{V}_{i}(G_{ij}\mbox{cos}\theta_{ij}+B_{ij}\mbox{sin}\theta_{ij}),     	
        \end{aligned}
		\label{Be3}\nonumber
		\end{equation}
where $\mathcal{H} \setminus i$ is the set of buses incident to the bus $i$.			

The measurement function for \textbf{reactive injection power} into the bus $i$:
		\begin{equation}
        \begin{aligned}
        h_{Q_{i}}(\cdot)  &= V_i \sum_{j \in \mathcal{H}} V_j(G_{ij}\sin\theta_{ij}-B_{ij}\cos\theta_{ij}),
        \end{aligned}
		\label{Ae2-2}\nonumber
		\end{equation}
with Jacobian expressions:
		\begin{equation}
        \begin{aligned}
     	\cfrac{\mathrm \partial{h_{Q_{i}}(\cdot)}}{\mathrm \partial \theta_{i}}&=
     	{V}_{i}\sum_{j \in \mathcal{H} \setminus i }
     	{V}_{j}(G_{ij}\mbox{cos}\theta_{ij}+B_{ij}\mbox{sin}\theta_{ij})\\
     	\cfrac{\mathrm \partial{h_{Q_{i}}(\cdot)}}{\mathrm \partial \theta_{j}}&=
     	{V}_{i}{V}_{j}(-G_{ij}\mbox{cos}\theta_{ij}-B_{ij}\mbox{sin}\theta_{ij})\\
     	\cfrac{\mathrm \partial{h_{Q_{i}}(\cdot)}}{\mathrm \partial V_{i}}&=
     	\sum_{j \in \mathcal{H} \setminus i }
     	{V}_{j}(G_{ij}\mbox{sni}\theta_{ij}-B_{ij}\mbox{cos}\theta_{ij})-2{V}_{i}B_{ii}	\\
     	\cfrac{\mathrm \partial{h_{Q_{i}}(\cdot)}}{\mathrm \partial V_{j}}&=
     	{V}_{i}(G_{ij}\mbox{sin}\theta_{ij}-B_{ij}\mbox{cos}\theta_{ij}).	     	
        \end{aligned}
		\label{Be3}\nonumber
		\end{equation}	

The measurement function for \textbf{current magnitude} at the branch connecting buses $i$ and $j$: 
		\begin{equation}
        \begin{gathered}
        h_{I_{ij}}(\cdot) = 
        [aV_i^2+bV_j^2-2V_iV_j(c\cos\theta_{ij}-d\sin\theta_{ij})]^{1/2}
        \end{gathered}
		\label{Ae3}\nonumber
		\end{equation}
		\begin{equation}
        \begin{aligned}
        a&=(g_{ij}+g_{si})^2+(b_{ij}+b_{si})^2;&
        b&=g_{ij}^2+b_{ij}^2\\
        c&=g_{ij}(g_{ij}+g_{si})+b_{ij}(b_{ij}+b_{si});&
        d&=g_{ij}b_{si}-b_{ij}g_{si}.
        \end{aligned}
		\label{Ae3}\nonumber
		\end{equation}
The Jacobian expressions corresponding to current magnitude measurement function $h_{I_{ij}}(\cdot)$ are as follows:	
		\begin{equation}
        \begin{aligned}
     	\cfrac{\mathrm \partial{h_{I_{ij}}(\cdot)}}{\mathrm \partial \theta_{i}}&=
     	\cfrac{V_iV_j(d\cos\theta_{ij}+c\sin\theta_{ij})}{I_{ij}}	\\
     	\cfrac{\mathrm \partial{h_{I_{ij}}(\cdot)}}{\mathrm \partial \theta_{j}}&=-
     	\cfrac{V_iV_j(d\cos\theta_{ij}+c\sin\theta_{ij})}{I_{ij}}\\
     	\cfrac{\mathrm \partial{h_{I_{ij}}(\cdot)}}{\mathrm \partial V_{i}}&=
     	\cfrac{V_j(d\sin\theta_{ij}-c\cos\theta_{ij})+aV_i}{I_{ij}}	\\
     	\cfrac{\mathrm \partial{h_{I_{ij}}(\cdot)}}{\mathrm \partial V_{j}}&=
     	\cfrac{V_i(d\sin\theta_{ij}-c\cos\theta_{ij})+bV_j}{I_{ij}}.      	
        \end{aligned}
		\label{Be5}\nonumber
		\end{equation}	

The Jacobian expressions corresponding to \textbf{voltage magnitude} and \textbf{voltage angle} measurement functions are as follows:	
		\begin{equation}
        \begin{aligned}
		\cfrac{\mathrm \partial{{V_{i}}}}{\mathrm \partial V_{i}}=1; & &
		\cfrac{\mathrm \partial{{\theta_{i}}}}{\mathrm \partial \theta_{i}}=1.
        \end{aligned}
		\label{Be6}\nonumber
		\end{equation}								
\bibliographystyle{IEEEtran}
\bibliography{smartgrid}

\begin{thebibliography}{10}
\providecommand{\url}[1]{#1}
\csname url@samestyle\endcsname
\providecommand{\newblock}{\relax}
\providecommand{\bibinfo}[2]{#2}
\providecommand{\BIBentrySTDinterwordspacing}{\spaceskip=0pt\relax}
\providecommand{\BIBentryALTinterwordstretchfactor}{4}
\providecommand{\BIBentryALTinterwordspacing}{\spaceskip=\fontdimen2\font plus
\BIBentryALTinterwordstretchfactor\fontdimen3\font minus
  \fontdimen4\font\relax}
\providecommand{\BIBforeignlanguage}[2]{{%
\expandafter\ifx\csname l@#1\endcsname\relax
\typeout{** WARNING: IEEEtran.bst: No hyphenation pattern has been}%
\typeout{** loaded for the language `#1'. Using the pattern for}%
\typeout{** the default language instead.}%
\else
\language=\csname l@#1\endcsname
\fi
#2}}
\providecommand{\BIBdecl}{\relax}
\BIBdecl

\bibitem{monticelli}
A.~Monticelli, \emph{State Estimation in Electric Power Systems: A Generalized
  Approach}, ser. Kluwer international series in engineering and computer
  science.\hskip 1em plus 0.5em minus 0.4em\relax Springer US, 1999.

\bibitem{abur}
A.~Abur and A.~Exp{\'o}sito, \emph{Power System State Estimation: Theory and
  Implementation}, ser. Power Engineering.\hskip 1em plus 0.5em minus
  0.4em\relax Taylor \& Francis, 2004.

\bibitem{wu}
F.~F. Wu, K.~Moslehi, and A.~Bose, ``Power system control centers: Past,
  present, and future,'' \emph{Proceedings of the IEEE}, vol.~93, no.~11, pp.
  1890--1908, Nov 2005.

\bibitem{pearl}
J.~Pearl, \emph{Probabilistic Reasoning in Intelligent Systems: Networks of
  Plausible Inference}.\hskip 1em plus 0.5em minus 0.4em\relax San Francisco,
  CA, USA: Morgan Kaufmann Publishers Inc., 1988.

\bibitem{bishop}
C.~M. Bishop, \emph{Pattern Recognition and Machine Learning}.\hskip 1em plus
  0.5em minus 0.4em\relax Springer, 2006.

\bibitem{loop}
Y.~Weiss and W.~T. Freeman, ``On the optimality of solutions of the max-product
  belief-propagation algorithm in arbitrary graphs,'' \emph{IEEE Transactions
  on Information Theory}, vol.~47, no.~2, pp. 736--744, Feb 2001.

\bibitem{kavcic}
Y.~Hu, A.~Kuh, A.~Kavcic, and D.~Nakafuji, ``Real-time state estimation on
  micro-grids,'' in \emph{Neural Networks (IJCNN), The 2011 International Joint
  Conference on}, July 2011, pp. 1378--1385.

\bibitem{ilic}
Y.~Weng, R.~Negi, and M.~Ilic, ``Graphical model for state estimation in
  electric power systems,'' in \emph{Smart Grid Communications, 2013 IEEE
  International Conference on}, Oct 2013, pp. 103--108.

\bibitem{schweppe}
F.~C. Schweppe and D.~B. Rom, ``Power system static-state estimation, part ii:
  Approximate model,'' \emph{IEEE Transactions on Power Apparatus and Systems},
  vol. PAS-89, no.~1, pp. 125--130, Jan 1970.

\bibitem{hansen}
P.~Hansen, V.~Pereyra, and G.~Scherer, ``Least squares data fitting with
  applications.''\hskip 1em plus 0.5em minus 0.4em\relax Johns Hopkins
  University Press, 2012, ch.~9, p. 166.

\bibitem{ping}
H.~A. Loeliger, J.~Dauwels, J.~Hu, S.~Korl, L.~Ping, and F.~R. Kschischang,
  ``The factor graph approach to model-based signal processing,'' \emph{Proc.
  of the IEEE}, vol.~95, no.~6, pp. 1295--1322, 2007.

\bibitem{jordan}
K.~P. Murphy, Y.~Weiss, and M.~I. Jordan, ``Loopy belief propagation for
  approximate inference: An empirical study,'' in \emph{Proc. of the 15th
  Conference on Uncertainty in Artificial Intelligence}, San Francisco, USA,
  1999, pp. 467--475.

\bibitem{extended}
M.~Cosovic and D.~Vukobratovic, ``State estimation in electric power systems
  using belief propagation: An extended {DC} model,'' in \emph{Signal
  Processing Advances in Wireless Communications SPAWC, 2016. 17th IEEE
  Workshop}, Edinburgh, United Kingdom, Jul. 2016.

\end{thebibliography}
\end{document}